\documentclass[pra,twocolumn,epsfig,nofootinbib,floatfix]{revtex4}
\usepackage{graphics}
\usepackage{epsfig}
\usepackage{amsfonts}
\usepackage{amsmath}
\usepackage{bm}

\usepackage{color}

\begin{document}
\title[Inverse population transfer of a repulsive Bose-Einstein condensate]
{Inverse population transfer of the repulsive Bose-Einstein
condensate in a double-well trap: strong interaction-induced
support}
\author{V.O. Nesterenko$^1$, A.N. Novikov$^1$ and E. Suraud$^2$}
\affiliation{$^1$ Bogoliubov Laboratory of Theoretical Physics,
Joint Institute for Nuclear Research, Dubna, Moscow region,
141980, Russia} \email{nester@theor.jinr.ru} \affiliation{$^2$
Laboratoire de Physique Quantique, Universit\'e Paul Sabatier, 118
Route de Narbonne, 31062 cedex, Toulouse, France}

\date{\today}

\begin{abstract}
An inverse population transfer of the repulsive Bose-Einstein
condensate (BEC) in a weakly bound double-well trap is explored
within the 3D time-dependent Gross-Pitaevskii equation. The model
avoids numerous common approximations (two-mode treatment,
time-space factorization,  etc) and closely follows the conditions
of Heidelberg experiments, thus providing a realistic description
of BEC dynamics. The transfer is driven by a time-dependent shift
of a barrier separating the left and right wells. It is shown that
completeness and robustness of the process considerably depend on
the amplitude and time profile of the shift velocity. Soft
profiles provide the most robust inversion. The repulsive
interaction substantially supports the transfer making it possible
i) in a wide velocity interval and ii)  three orders of magnitude
faster than in the ideal BEC.
\end{abstract}

\maketitle

\section{Introduction}

The population inversion is a typical problem met in various
branches of physics (ultracold gases and condensates
\cite{Gati_07,Torr_12_rev,Yuk}, atomic and molecular physics
\cite{Kral}, etc.). The problem is easily solvable, if it is
linear and accepts an adiabatic evolution, see e.g. the
Landau-Zener scenario \cite{Lan,Zen}. However, if there are
significant nonlinear effects or/and we need a rapid but robust
transfer, the problem becomes nontrivial, e.g. for a transport of
Bose-Einstein condensate (BEC) \cite{Nest_LP_09,Nest_JPB_09}. The
inverse population transfer of a repulsive BEC in a double-well
trap is the relevant example of such a complex
problem\cite{Nest_JPB_09}: the repulsive interaction in BEC leads
to a strong nonlinearity and a limited life-time of BEC requests a
rapid transfer. However, a rapid process, if not specially
designed, is usually spoiled by dipole oscillations at the final
state. The question is how to produce a robust (without final
dipole oscillations) and rapid {\it nonlinear} population
inversion (NLPI) in this case?  Another important aspect is how
does the nonlinearity affect the process?

In principle, this problem is a subject of so-called
shortcuts-to-adiabaticity (StA) methods which have made a
significant progress over the last years, see \cite{Torr_12_rev}
for an extensive review. However, to our knowledge, these methods
have not yet been derived for our particular case: NLPI for BEC in
a double-well trap. Moreover, even though some StA methods, like
the optimal control theory \cite{oct_Gross,oct_Brief},  may be
potentially implemented in this case, their protocols might be too
complicated and parameter-sensitive to be realized in experiment,
while we need a simple prescription with a minimal number of
control parameters. We will show that such a prescription can be
built using the setup  of Heidelberg experiments \cite{Gati_07},
where a time-dependent shift $x_0(t)$ of the barrier is used as a
suitable control parameter driving the trap asymmetry and thus the
population transfer.

In the present study, the barrier shift is used to produce NLPI of
BEC in a double-well trap. The calculations are performed within
the three-dimensional (3D) time-dependent Gross-Pitaevskii
equation \cite{GPE} for the total order parameter covering both
left and right parts of the condensate. Our model
\cite{Nest_JPB_12} is free from  numerous approximations (two-mode
treatment, time-space factorization of the order parameter, etc)
widely used in investigation of BEC dynamics in a double-well trap
(see e.g. \cite{Nest_JPB_09} and references therein) and closely
follows prescriptions of the Heidelberg's experiments
\cite{Albiez_exp_PRL_05,Gati_APB_06}, thus providing quite
realistic picture.

The NLPI is explored for different magnitudes and time profiles of
the transfer velocity, thus covering  adiabatic and rapid
scenarios. Both ideal and repulsive BECs are considered to
estimate the nonlinear effect caused by the interaction between
BEC atoms. As shown below, the repulsive interaction strongly
favours the NLPI. This is in agreement with our previous results
obtained within less involved model \cite{Nest_JPB_09}.

The paper is organized as follows.  The calculation scheme is
sketched in Sec. \ref{sec:calc_scheme}, the results are discussed
in Sec. \ref{sec:results}, the summary is done in Sec.
\ref{sec:summary}.

\section{Calculation scheme}
\label{sec:calc_scheme}

 We use the 3D time-dependent Gross-Pitaevskii equation (GPE) \cite{GPE}
\begin{equation}\label{GPE}
 i\hbar\frac{\partial\Psi}{\partial t}({\bf r},t) =
[-\frac{\hbar^2}{2m}\nabla^2 + V({\bf{r}},t) + g_0|\Psi({\bf
r},t)|^2]\Psi({\bf r},t)
\end{equation}
for the total order parameter $\Psi({\bf r},t)$ describing the BEC
in both left and right wells of the trap. Here $g_0=4\pi\hbar^2
a_s /m$ is the interaction parameter, $a_s$ is the scattering
length, and $m$ is the atomic mass. The trap potential
\begin{eqnarray} \label{trap_pot}
V({\bf{r}},t)&=& V_{\rm{con}}({\bf{r}})+V_{\rm{bar}}(x,t)
 \\
 &=&\frac{m}{2}(\omega^2_x x^2+\omega^2_y
y^2+\omega^2_z z^2) \nonumber
\\&+& V_0 \cos^2(\pi (x-x_0(t))/q_0) \nonumber
\end{eqnarray}
includes the anisotropic harmonic confinement and the barrier in
$x$-direction, whose position is driven by the control parameter
$x_0(t)$. Furthermore, $V_0$ is the barrier height and $q_0$
determines the barrier width.

Following the Heidelberg experiment
\cite{Albiez_exp_PRL_05,Gati_APB_06} for measuring Josephson
oscillations (JO) and macroscopic quantum self-trapping (MQST), we
consider the BEC of N=1000 $^{87}$Rb atoms with $a_s=5.75$ nm. The
trap frequencies are $\omega_x=2\pi\times 78$ Hz,
$\omega_y=2\pi\times 66$ Hz, $\omega_z=2\pi\times 90$ Hz, i.e.
$\omega_y+\omega_z=2\omega_x$. The barrier parameters are
$V_0=420\times h$ Hz and  $q_0=5.2 \; \mu$m.
For the symmetric trap ($x_0$(t)=0), the distance between the
centers of the left and right wells is $d=$4.4 \;$\mu$m. This
setup has been previously used in our exploration of JO/MQST in
weak and strong coupling regimes \cite{Nest_JPB_12}.

The static solutions of GPE are found within the damped gradient
method \cite{DGM} while the time evolution is computed within the
time-splitting  \cite{TSM} and fast Fourier-transformation
techniques. The total order parameter $\Psi({\bf r},t)$ is
determined in a 3D cartesian grid. The requirement
$\int^{-\infty}_{+\infty}dr^3 |\Psi({\bf r},t)|^2 =N$ in time is
directly fulfilled by using an explicit unitary propagator.
Reflecting boundary conditions are used, though they have no
impact on the dynamics because the harmonic confinement makes them
effectless. No  time-space factorization of the order parameter is
used. The conservation of the total energy $E$ and complete number
of atoms $N$ is perfectly controlled.

The populations of the left (L) and right (R) wells are computed
as
\begin{equation}\label{N_LR}
N_{j}(t)=\int^{+\infty}_{-\infty}dr^3 |\Psi_{j}({\bf r},t)|^2,
\end{equation}
with $j = L, R$, $\Psi_{L}({\bf r},t)=\Psi(x\le 0,y,z,t)$,
$\Psi_{R}({\bf r},t)=\Psi(x\ge 0,y,z,t)$ and $N_L(t)+N_R(t)=N$.
The normalized population imbalance is
$z(t) = (N_L(t)-N_R(t))/N$.

The population inversion means that BEC population characterized
at the initial time t=0 by $N_L(0) > N_R(0)$ is changed during the
time interval T to the inverse population $N_L(T) < N_R(T)$
where $N_L(T)=N_R(0)$ and  $N_R(T)=N_L(0)$.

Following the technique of \cite{Albiez_exp_PRL_05,Gati_APB_06},
the initial stationary asymmetric BEC state is produced by keeping
the barrier right-shifted from the symmetric case ($x(0) > 0$).
The value of the shift is adjusted to provide the given initial
populations $N_L(0)$ and $N_R(0)$. The population inversion is
generated by the time-dependent left shift of the barrier from
$x(0)$ to $x(T)=-x(0)$ with the velocity $v(t)$. Thus asymmetry of
the double-well trap is changed to the opposite one.

The quality of the inversion is characterized by its completeness
$P=-z(T)/z(0)$ (the ratio of the final and initial population
imbalance) and noise $n=A_d/N$ where $A_d$ is amplitude of dipole
oscillations in the final state, i.e.
$A_d=\rm{max}\{N_{L,R}\}-\rm{min}\{N_{L,R}\}$ for $t>T$.

Two velocity time profiles are used: i) the sharp rectangular one
with the constant $v_c(t)=v^c_0$ at $0 < t < T$ and $v_c(t) =$ 0
beyond the transfer time and ii) the soft one $v_s(t)=v^s_0
\cos^2(\frac{\pi}{2}+\frac{\pi t}{T})$ with $v_s(0)=v_s(T) \sim 0$
and  $v_s(T/2)=v^s_0$. For the total barrier shift $D=2 x(0)$ in
the inversion process of duration T, the velocity amplitudes are
$v^c_0=v^s_0=D/T$. The profile $v_c(t)$ is simple. However, it
sharply changes from 0 to $v^c_0$ at t=0 and T and so is not
adiabatic. The second profile $v_s(t)$ is more complicated but
softer and thus closer to an adiabatic case.

The transfer time T has natural lower and upper limits. It cannot
be longer than the BEC lifetime ($\sim$ 3 sec). Also it cannot be
too short since then the transfer would be too sharp and cause in
the final state large undesirable dipole oscillations (see
discussion in the next section). The same reasons determine the
upper and lower limits for the transfer velocities.

\section{Results and discussion}
\label{sec:results}

Figure 1 exhibits  the trap potential in x-direction,
\begin{equation}
V_x(x,t)=\frac{m}{2}\omega^2_x x^2 + V_0 \cos^2(\pi
(x-x_0(t))/q_0) ,
\end{equation}
 calculated for the initial t=0, intermediate t=T/2 and final t=T
times of the inversion process driven by the barrier shift
$x_0(t)$ with t$\in$[0,T]. For the same times, the BEC density in
x-direction,
\begin{equation}
\rho (x,t)=\int^{+\infty}_{-\infty} dy dz |\Psi (x,y,z,t)|^2 ,
\end{equation}
obtained for an adiabatic inversion of a long duration T is shown.
The ideal and repulsive BECs with N=1000 atoms are considered.
Following the plots a) and d), the initial populations of the left
and right wells are $N_L(0)$=800 and $N_R(0)$=200, i.e. with the
initial population imbalance  $z(0)$=0.6.    An adiabatic
evolution provides a robust population inversion to final
$N_L(T)$=200, $N_R(T)$=800 and  $z(T)$=-0.6. At the intermediate
time t=T/2, the trap and populations are symmetric. The initial
state is stationary by construction. The intermediate and final
states, if obtained adiabatically, may be also treated as
stationary.
\begin{figure*}
\begin{center}
\includegraphics[width=11cm]{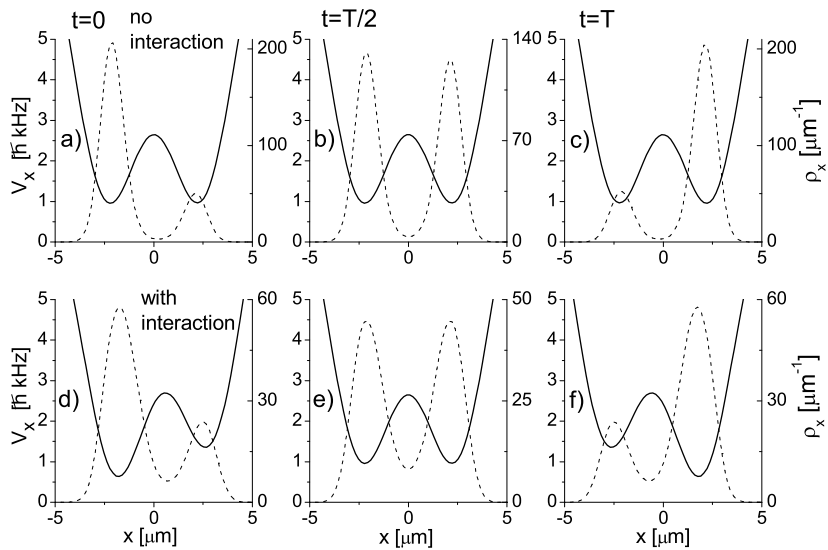}
\end{center}
\caption{The double-well trap potential $V_x(x)$ (bold curve) and
BEC density $\rho_x(x)$ (dash curve) at initial (t=0),
intermediate (t=T/2) and final inverse (t=T) states of the
adiabatic inversion, calculated without (upper plots) and with
(bottom plots) the repulsive interaction between BEC atoms. In
both cases, the initial populations of the left and right wells
are $N_L(0)$=800 and $N_R(0)$=200.}
\end{figure*}
\begin{figure}[h]
\begin{center}
\includegraphics[width=8cm]{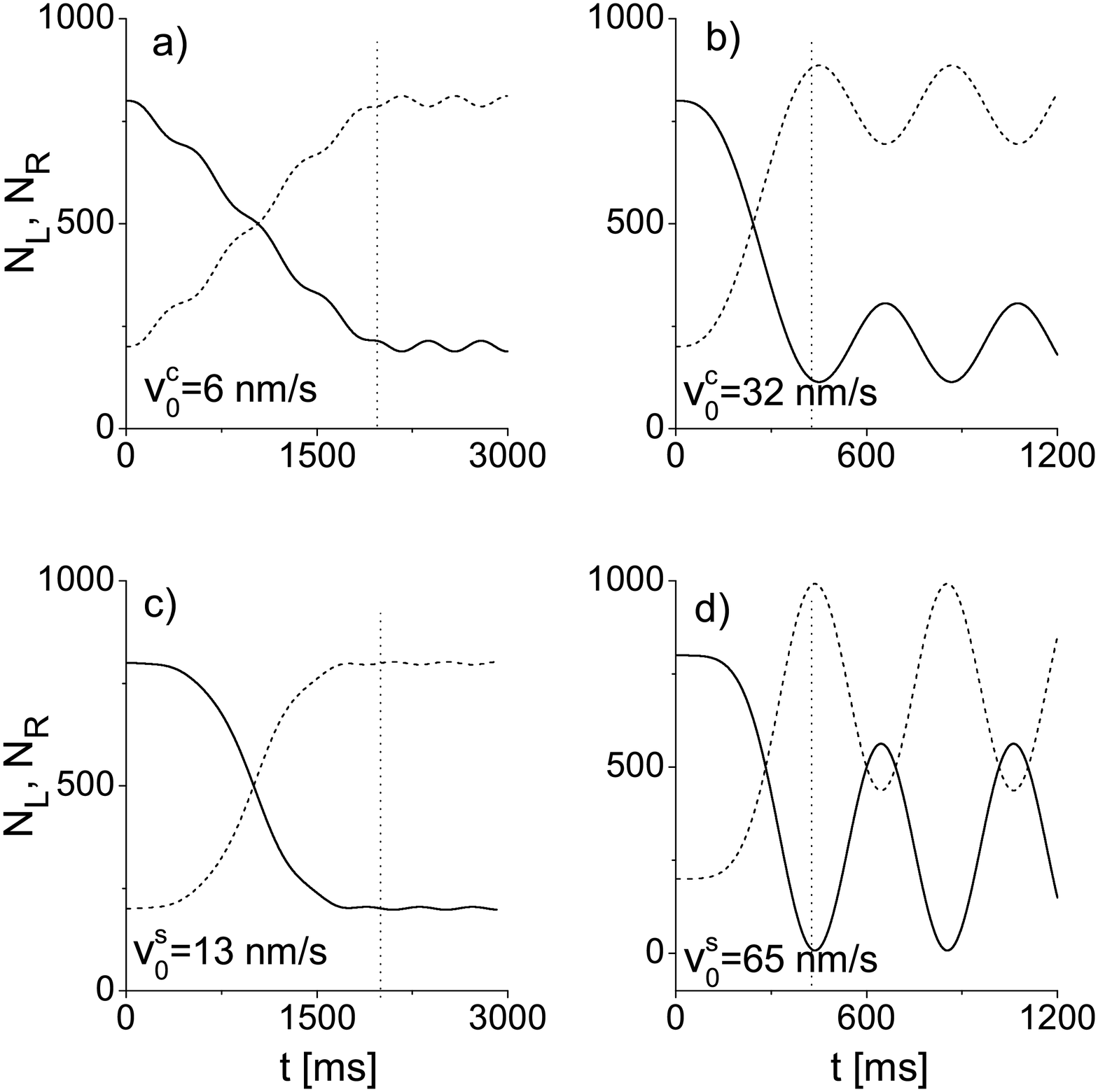}
\end{center}
\caption{Ideal (without the interaction) BEC. Time-dependent
evolution ofpopulations $N_L(t)$ (solid curve) and $N_R(t)$ (dash
curve), calculated for the initial conditions $N_L$(0)=800 and
$N_R$(0)=200, i.e. $z$(0)=0.6. Durations of the barrier shift
($T$=1.7 $\mu$s in a),c) and $T$=0.45 $\mu$s in b),d)) are
indicated by vertical dotted lines. The transfers with the
constant (rectangular) $v^c_0$ (upper plots) and soft
$v_s(t)=v^s_0 cos^2(\frac{\pi}{2}+\frac{\pi t}{T})$ (bottom plots)
velocities of the barrier shift are considered. In every plot, the
velocity amplitudes are depicted.}
\end{figure}
\begin{figure}[h]
\begin{center}
\includegraphics[width=8cm]{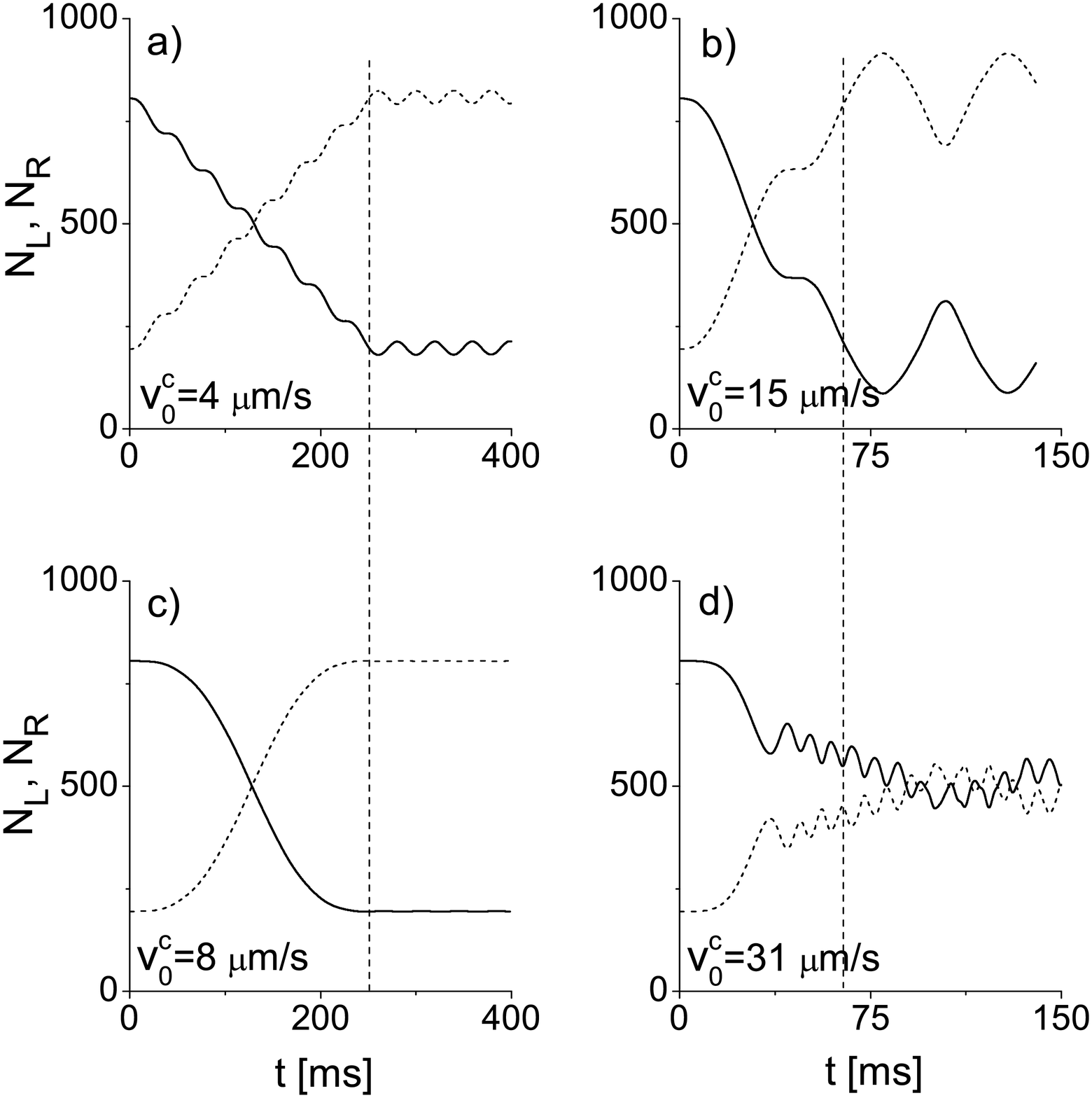}
\end{center}
\caption{The same as in Fig. 2 but for repulsive BEC.}
\end{figure}

Upper plots of Fig. 1 show that for getting  the initial
$z(0)$=0.6 in the ideal BEC, a small trap asymmetry with
$x_0(0)$=0.003 $\mu$m is sufficient. The overlap of the left and
right parts of the condensate at the center of the trap is very
small and corresponds to the case of a weak coupling. The energy
difference  between the ground and first excited states at the mid
of the transfer  (plot b)) is $\Delta E(T/2)$ = 0.005 $\hbar$ kHz.
Such a tiny value confirms that the coupling and corresponding
barrier penetrability are indeed small.

For the repulsive BEC (bottom plots), the initial  $N_L(0)$=800
and $N_R(0)$=200 are obtained at much larger asymmetry with
$x_0(0)$=0.5 $\mu$m. The energy splitting $\Delta E(T/2)$ reaches
0.036 $\hbar$ kHz. This is because the repulsive interaction
significantly increases the chemical potential $\mu$ of the system
and thus the coupling between the left and right parts of BEC.
Then, to get the initial {\it stationary} population imbalance
$z(0)$=0.6, one should weaken the coupling by considerably
increasing  the asymmetry. As compared to the ideal BEC, the
repulsive condensate has wider density bumps with much stronger
overlap at the center of the trap. The coupling between the left
and right BEC parts is not yet weak anymore. Nevertheless, the
NLPI described below has occurred through tunneling.

Some examples of the time evolution of the populations
$N_{L,R}(t)$ in the ideal BEC (no interaction) are given in Fig.
2. The evolution is driven by the barrier shift with the
rectangular $v_c(t)$ (upper plot) and soft  $v_s(t)$ (bottom
plots) velocitiy profiles. The total shift is $D=2 x(0)=$6 nm. For
each example, the velocity amplitudes $v^c_0$ and $v^s_0$ obtained
for a given transfer duration $T$ are indicated. It is seen that,
at low velocities (plots a),c)) corresponding to a long duration
T=1.8 $\mu$s, we get rather robust population inversion. This
transfer is close to an adiabatic evolution. The final state is
about stationary for $v_s(t)$ and somewhat spoiled by dipole
oscillations for $v_c(t)$. The latter is caused by the sharp
change of $v_c(t)$ from 0 to $v_c(0)$ and back at the beginning
and end of the process. In this sense, the $v_s(t)$-transfer  is
much softer and thus  more adiabatic. Fig. 2 also shows that the
barrier-shift technique may be used not only for the population
inversion  (plots (a,c)) but also for production of MQST (plot
(b)) and JO (plot (d))  supplementing the process. In our task,
the JO/MQST come as undesirable dipole oscillations.

In Fig. 3,  similar examples are presented for the repulsive BEC.
At first glance, the non-linear evolution resembles the linear one
given in Fig. 2 except for  plot d)) where the final state
converges to a symmetric form with $N_L(T) \sim N_R(T) \approx
500$ or $z(T)\approx $0). Like in the linear case,  a slow
transfer (plots a),c)) results in a robust NLPI while a faster
process (plots b),d)) spoils the final state by dipole
oscillations (plot b)) or even breaks the inversion at all (plot
d)). However, the nonlinearity drastically changes rates of the
process. Now the robust NLPI can be produced at much shorter time
(T=250 $\mu$s instead of T=1800 $\mu$s for ideal BEC) and much
faster velocities ($\mu$/s instead of nm/s). So, despite the NLPI
requires much stronger asymmetry and longer barrier shift (1
$\mu$m against 0.006 $\mu$m for the ideal BEC), the process
becomes much faster. Namely, the velocities become about three
order of magnitude higher (!), i.e. far beyond the adiabatic case.
Therefore the repulsive interaction greatly favours the population
inversion and this effect is indeed huge. The  reason of the
effect is simple. As mentioned above, the repulsive interaction
significantly enhances the chemical potential $\mu$, which in turn
results in a dramatic increase of the barrier penetrability. The
coupling between BEC fractions becomes strong and the inversion is
realized much faster.
\begin{figure}[h]
\begin{center}
\includegraphics[width=8cm]{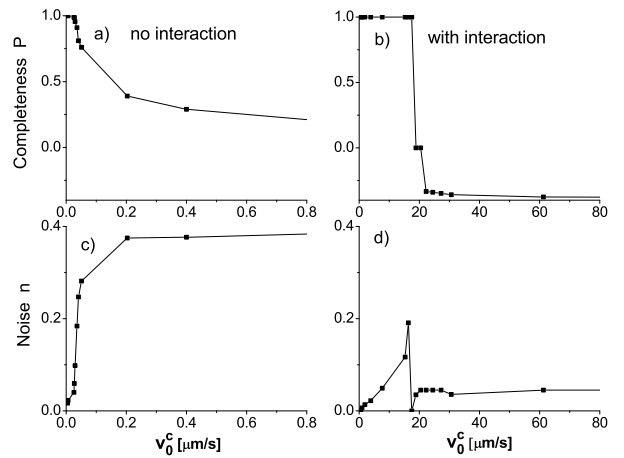}
\end{center}
\caption{Completeness a)-b) and noise c)-d) of the population
inversion for BEC without (left plots) and with (right plots)
repulsive interaction versus the amplitude $v^c_0$ of the velocity
$v_c(t)$. The initial population imbalance is $z$(0)=0.6.}
\end{figure}
\begin{figure}[h]
\begin{center}
\includegraphics[width=8cm]{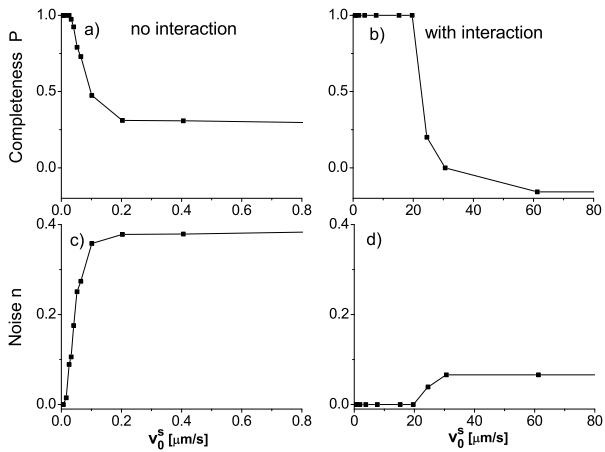}
\end{center}
\caption{The same as in Fig. 4 but for velocity profile $v_s(t)$.}
\end{figure}

A more general information on the robustness of the population
inversion is presented in Figs. 4 and 5 where the completeness $P$
and noise $n$ of the inversion are illustrated for a wide range of
velocity amplitudes. The inversions for the ideal and repulsive
BEC are compared. In Fig. 4, the sharp velocity profile $v_c(t)$
is used. Following the plots a) ,c) for the ideal BEC, a complete
inversion ($P$=1) takes place only at a small velocity $v^c_0 <$
0.04 $\mu$m/s. The inversion is somewhat spoiled by a weak noise
$n=$ 0.02 - 0.04. The smaller the velocity, the weaker the noise.
For $v^c_0 >$ 0.04 $\mu$m/s, we see a gradual destruction of the
inversion, accompanied by an enhanced noise. For even larger
velocities, the inversion breaks down ($P \to $ 0) and the final
state is characterized by strong Rabi oscillations ($n \to$ 0.4).
The latter effect is caused by the instant change of the process
velocity from zero to $v^c_0$ at t=0 and back at t=T.

Following Fig. 4 (b),d)), inclusion of the repulsive interaction
drastically changes the results. There appears a wide plateau, $0
< v^c_0 \le$ 19 $\mu$m/s, with about complete inversion $P \approx
$ 1. The repulsive interaction thus allows to get the inversion in
a much wider velocity interval and, what is important, about three
order of magnitude (!) faster than for the ideal BEC. Following
our estimations, this is mainly caused by a considerable increase
of the chemical potential $\mu$, caused by the repulsive
interaction, and thus  increasing the barrier penetrability. Note
that the nonlinearity plays here an important but auxiliary role,
which is reduced to a mechanism of rising the chemical potential.
The net effect should depend on the form of the barrier. It should
be strong for barriers whose penetrability increases with the
excitation energy (e.g. Gaussian and $\cos^2(\pi (x-x_0(t))/q_0)$
 barrier shapes) and suppressed for barriers with an
energy-independent penetrability (e.g. rectangular shape).

The plot Fig. 4 b) exhibits a noise (Rabi oscillations at the
final state) in both inversion $v^c_0 \le$ 19 $\mu$m/s and beyond
$v^c_0 \ge$ 19 $\mu$m/s regions. In the former region, the noise
rises with the velocity, i.e. the faster the process, the less
robust the process. At $v^c_0 \ge$ 19 $\mu$m/s, the inversion
breaks down. Unlike the linear case,  the transfer completeness
$P$ does not tend to zero but to the negative value $P \approx
$-0.7. This means that $z(0)$ and $z(T)$ have the same sign, i.e.
the process results only in a modest population transfer, keeping
the initial inequality $N_L > N_R$  at t=T.

In Figure 5, the similar analysis is done for the softer (more
adiabatic) velocity profile $v_s(t)$. The results are very similar
to those in Fig. 4. The only but important difference is that, in
the repulsive BEC, the inversion at $v^s_0 \le$ 20 $\mu$m/s is
accompanied by much less noise as compared to the previous
$v_c(t)$ case (see plot d)). So, as might be expected, the softer
(and thus more adiabatic) velocity profile $v_s(t)$ provides a
much better inversion than the sharp profile $v_c(t)$.

The physical sense of the critical velocity $v_{\rm{crit}}
\approx$ 19-20 $\mu$m/s which marks  the break of inversion for
both $v_c(t)$ and $v_s(t)$ regimes should be clarified. Following
our analysis, $v_{\rm{crit}}$ does not corresponds to the
destruction of adiabaticity (indeed we have about the same
$v_{\rm{crit}}$ for less and more adiabatic profiles $v_c(t)$ and
$v_s(t)$),  but is rather determined by the transfer capacity
defined as the multiplicative effect of the barrier penetrability
$w$ and transfer duration $T$. Just these two values suffice to
control the number of atoms to be transferred. Time $T$ can be
enough ($v^{c,s}_0 < v_{\rm{crit}}$) or not ($v^{c,s}_0 >
v_{\rm{crit}}$) for the complete inversion. For the repulsive BEC,
the barrier penetrability $w$ is high and so the full inversion
can be fast which explains the high $v_{\rm{crit}}$ and wide NLPI
plateau in Figs. 4 b) and 5 b). In the ideal BEC, the
penetrability $w$ is much weaker and thus longer times (lower
velocities) are necessary for the inversion. Altogether, we see a
strong support of the inversion by the repulsive interaction. The
nonlinearity does not destroy but instead greatly favours the
inversion, making it much faster.

These findings are in accordance with  our previous results for
the complete transport of BEC from the left to the right well,
obtained within the simplified model employing the two-mode
approximation \cite{Nest_JPB_09}. In that study,  the appearance
of a wide velocity plateau due to the repulsive interaction was
also observed. Note that velocity of the process has also a lower
limit (e.g. caused by the finite lifetime of BEC, which is
commonly a few seconds).

\section{Summary}
\label{sec:summary}

The complete population inversion of the repulsive BEC in a
double-well trap was investigated within the time-dependent
three-dimensional Gross-Pitaevskii equation, following parameters
of experiments of the Heidelberg group
\cite{Albiez_exp_PRL_05,Gati_APB_06}. The calculations are
performed beyond usual  approximations (two-mode, etc) in the
description of tunneling and transport dynamics. The inversion is
driven by a time-dependent barrier shift performed with different
velocity regimes. As might be expected, a soft velocity profile
$v(t)$ gives a more robust inversion than the sharp one.

The most remarkable result is a significant support of the
complete inversion by the repulsive interaction between BEC atoms.
Due to the interaction (and related nonlinearity of the problem),
the inversion can be produced in a wide velocity interval.
Moreover, the process can be three orders of magnitude (!) faster
than in the ideal BEC. Thus the transfer can be done far beyond
the adiabatic requirements. These results are in accordance with
our previous findings obtained within the two-mode approximation
approach \cite{Nest_JPB_09}. The interaction effect is mainly
reduced to the rise of chemical potential. Hence it should depend
on the barrier form, being strong for barriers whose penetrability
increases with the excitation energy and suppressed for barriers
with energy-independent penetrability.

\section*{Acknowledgments}
The work was partly supported by the RFBR grant  11-02-00086à and
Institut Universitaire de France. We thank Prof. D.
Gu´\'ery-Odelin for useful discussions.

\end{document}